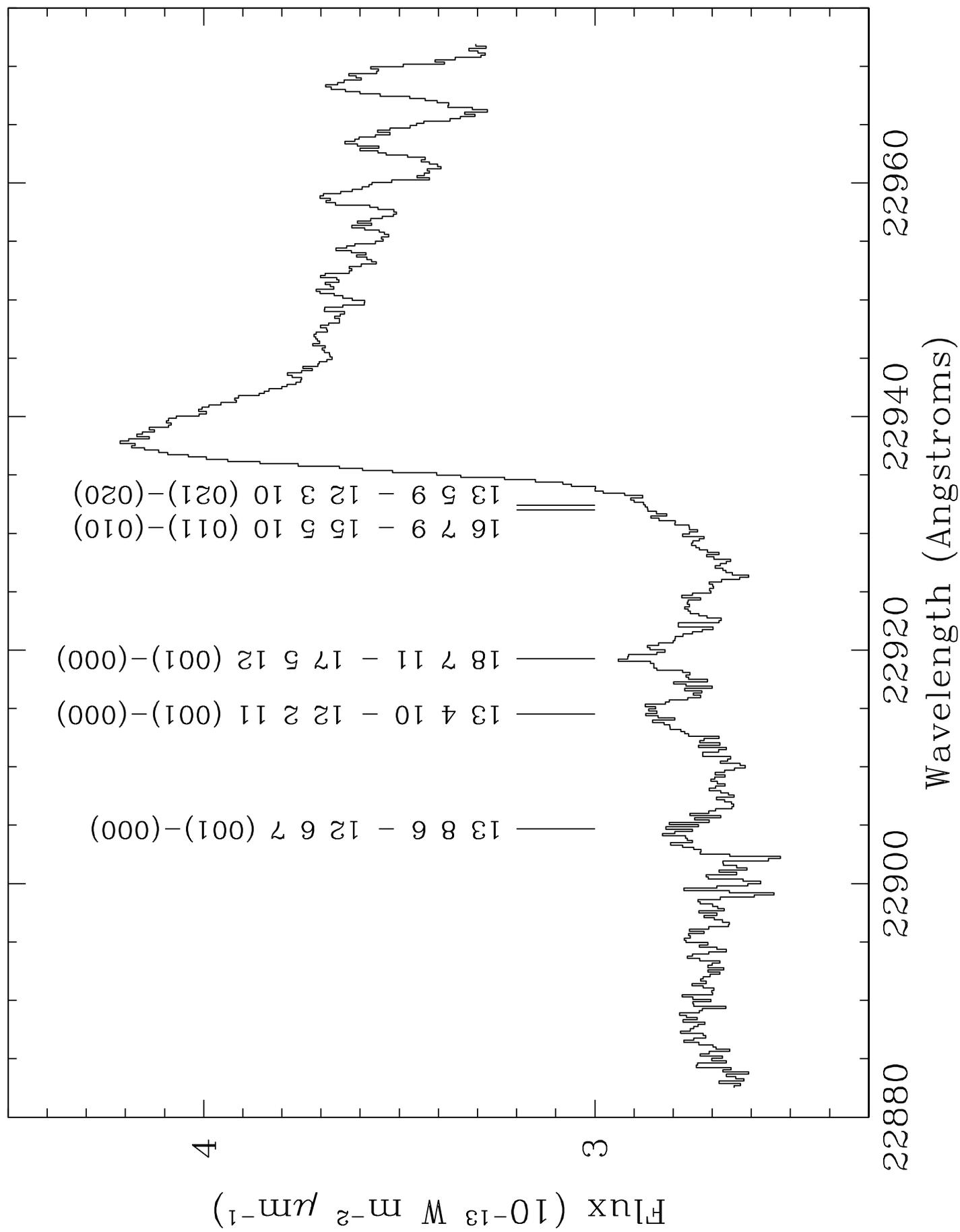

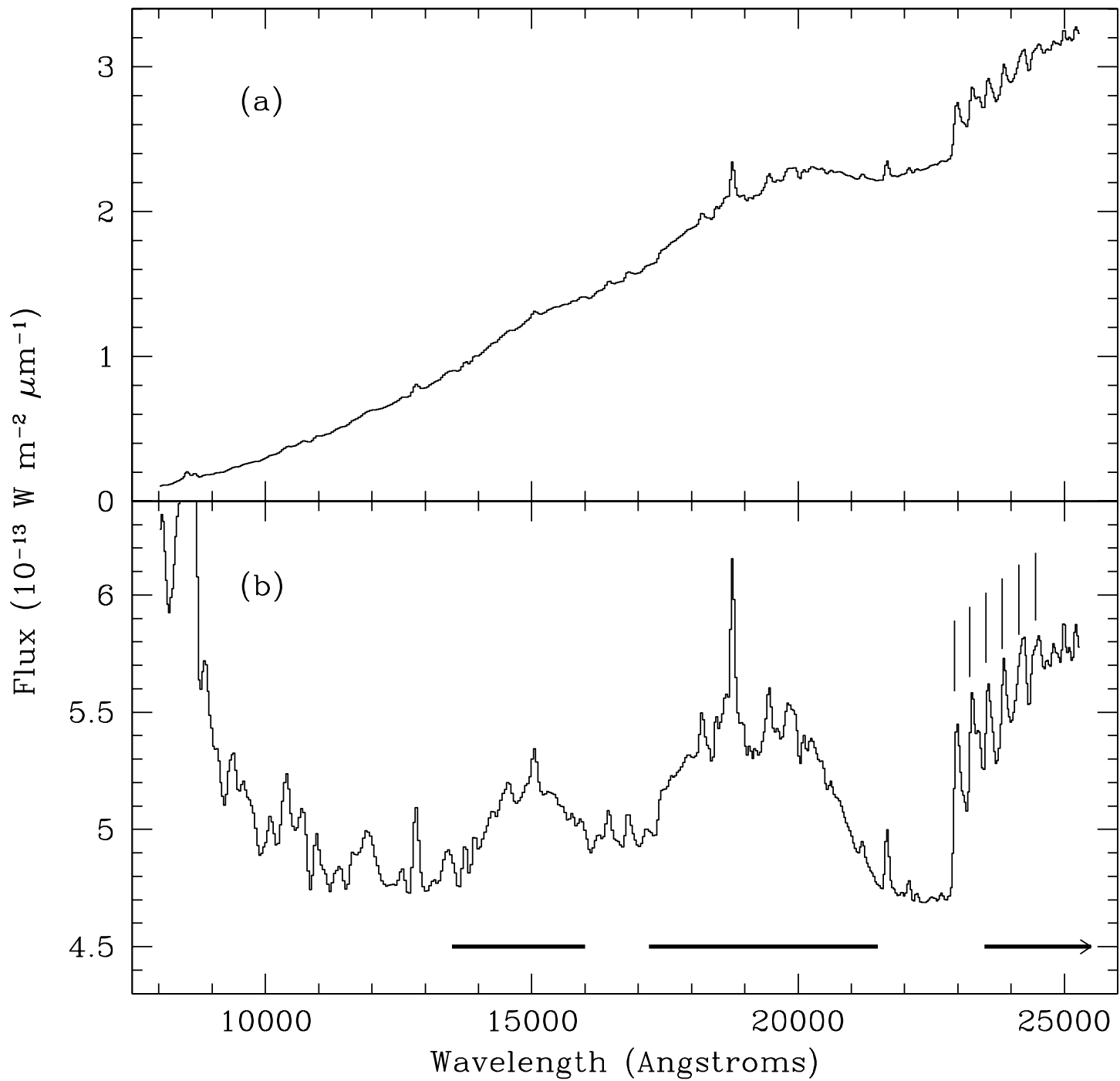

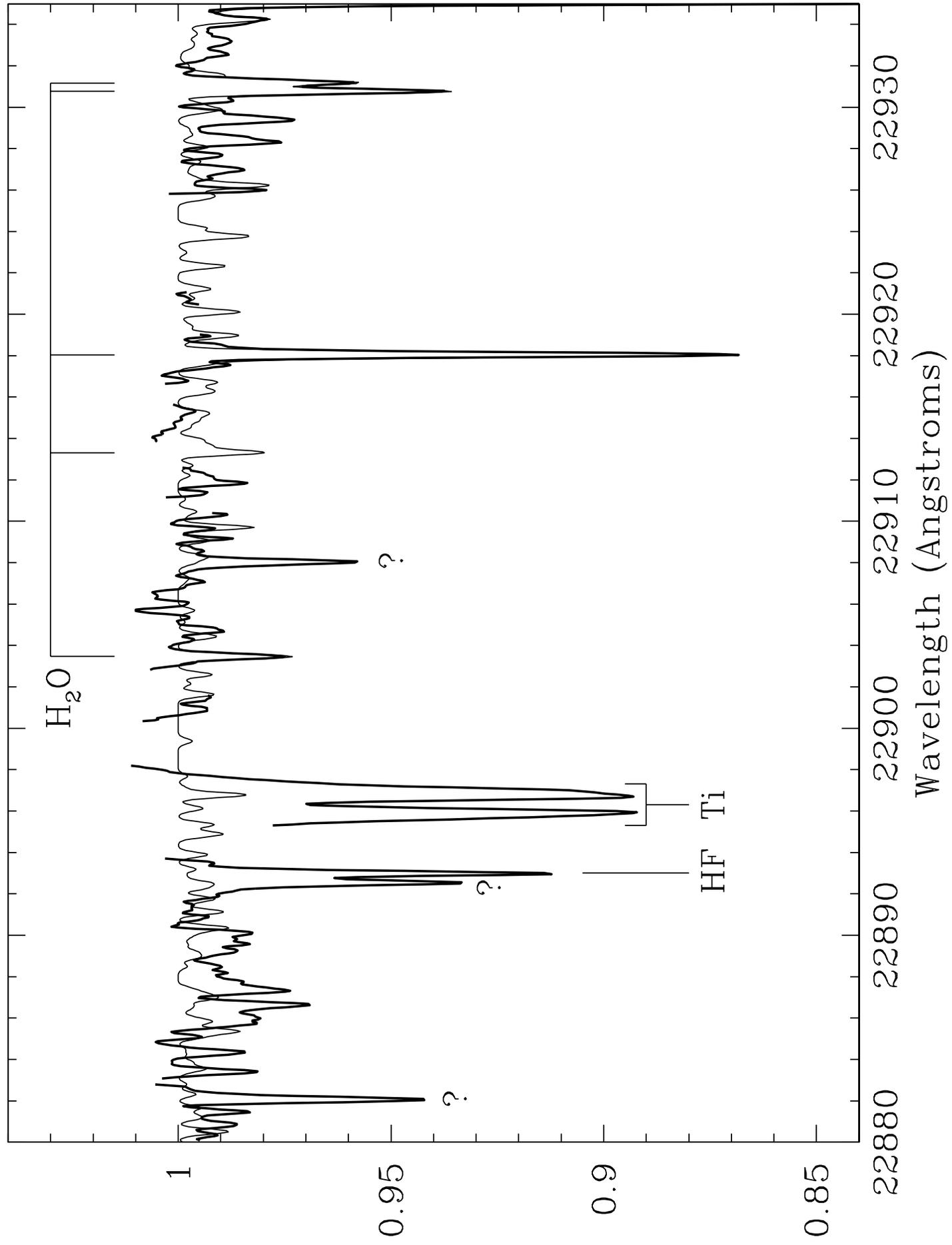

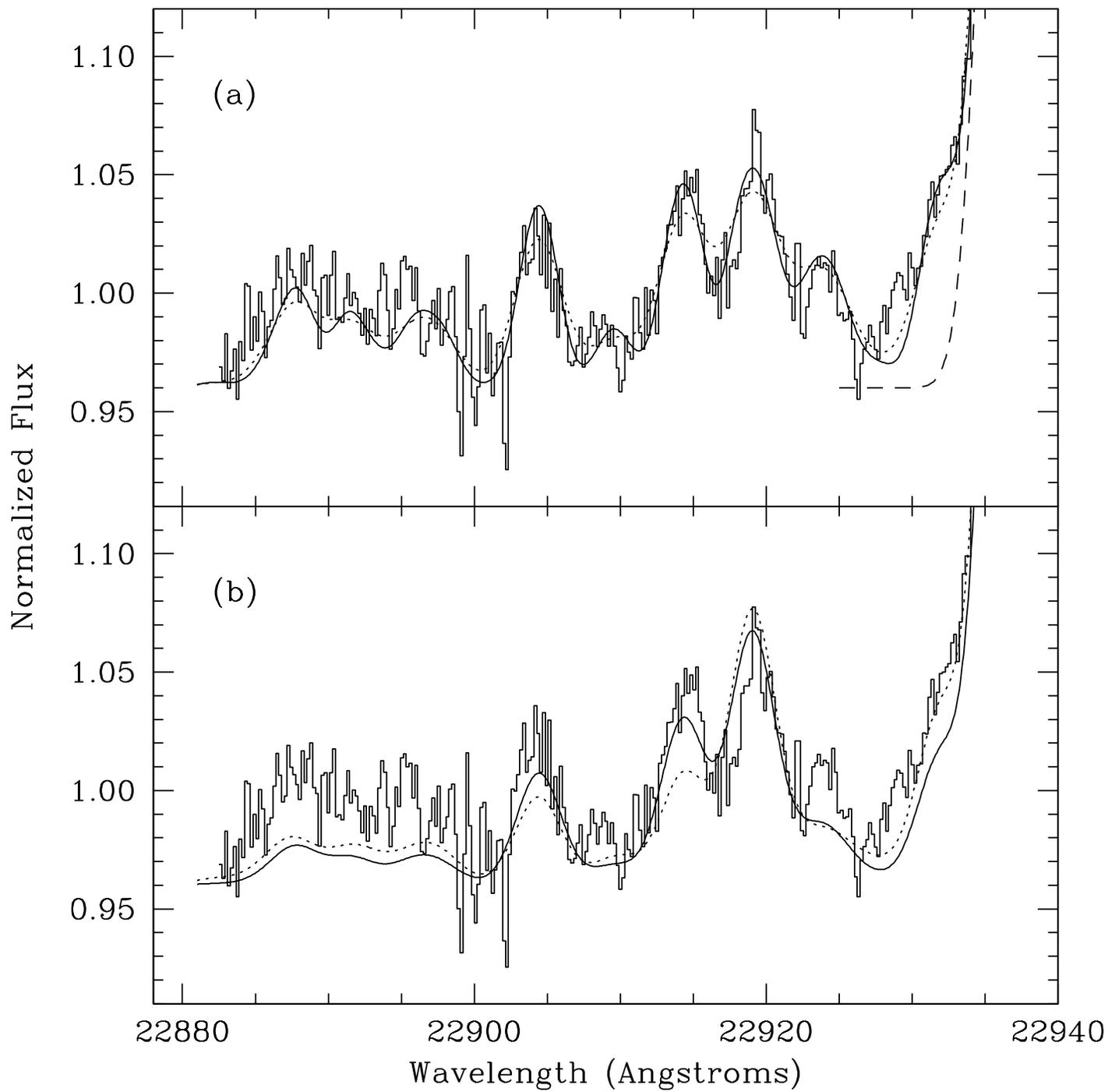

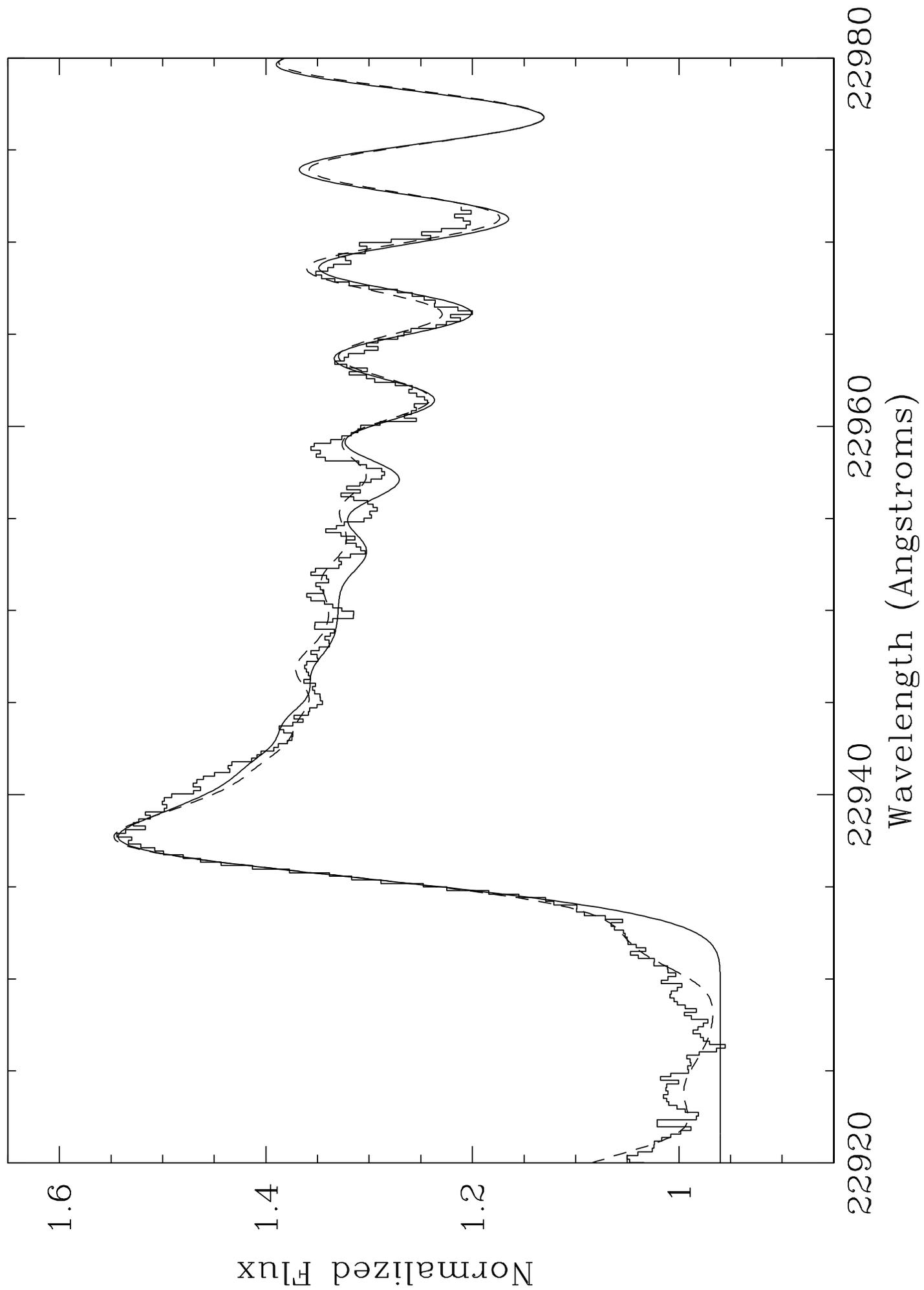

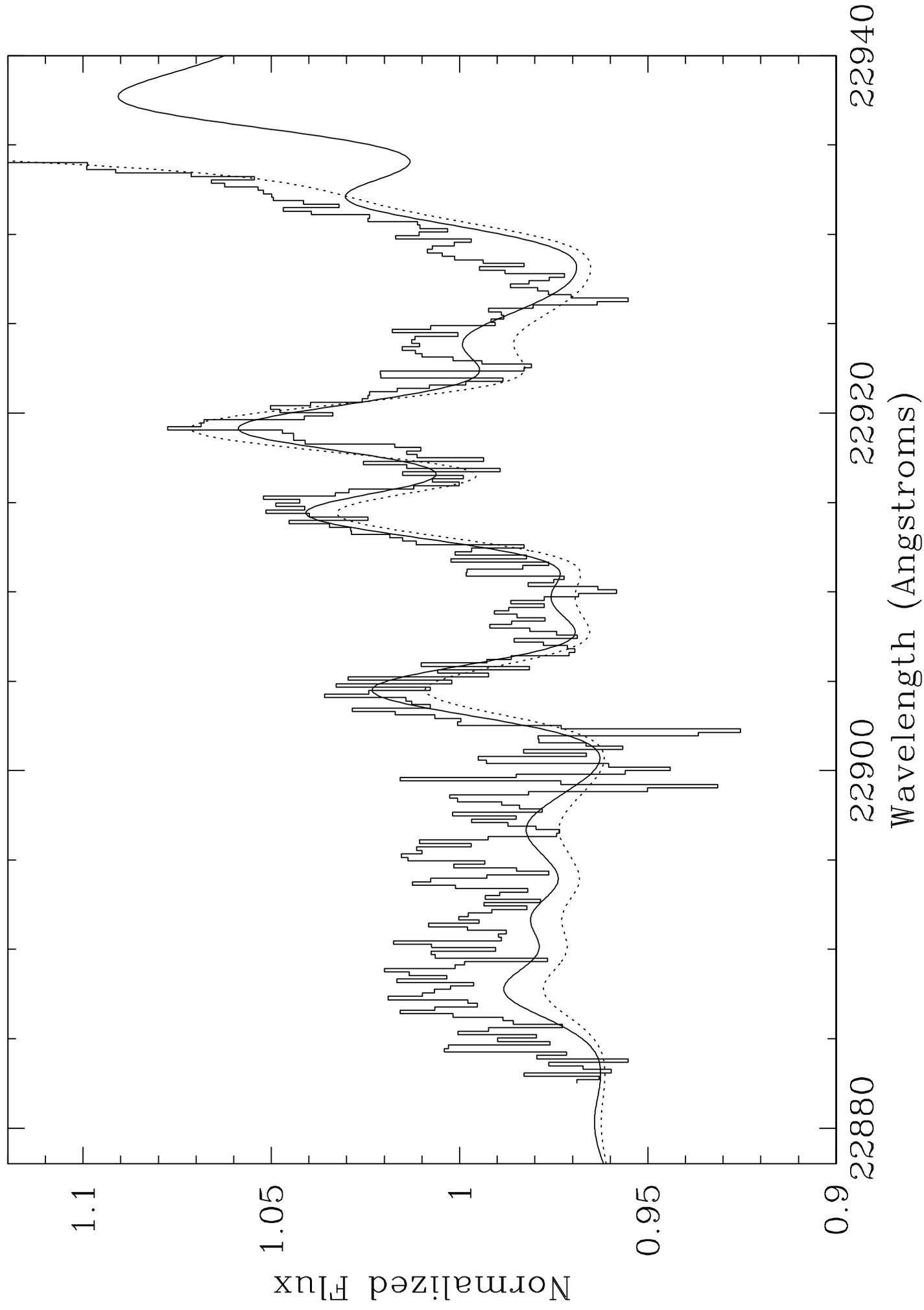

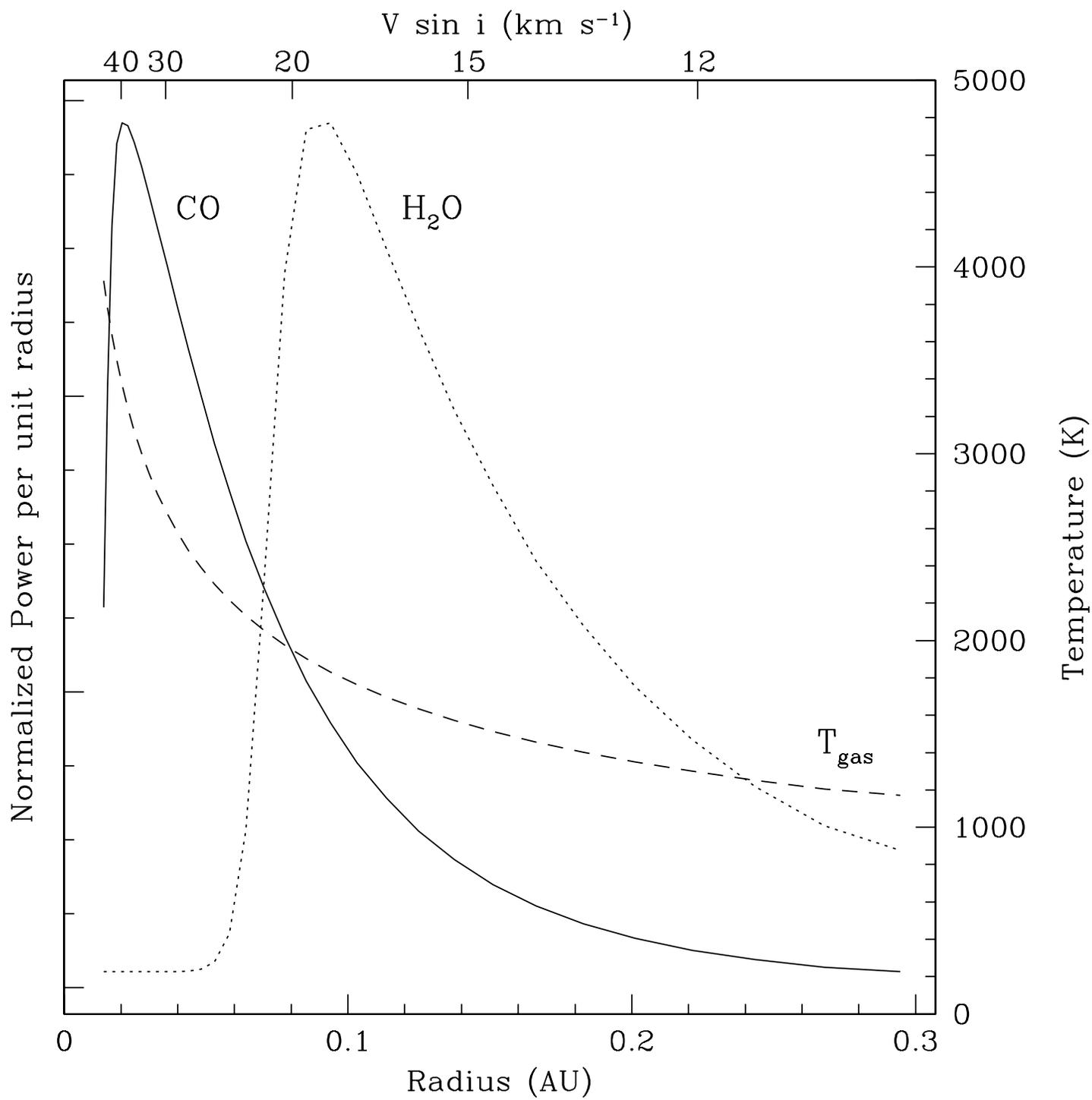

# Hot $H_2O$ Emission and Evidence for Turbulence

# in the Disk of a Young Star

John S. Carr[1]

Naval Research Laboratory, Code 7213, Washington, D. C. 20375; carr@nrl.navy.mil

Alan T. Tokunaga[1]

Institute for Astronomy, University of Hawaii, 2680 Woodlawn Drive, Honolulu, HI 96822;

tokunaga@ifa.hawaii.edu

Joan Najita[1]

National Optical Astronomy Observatory, 950 North Cherry Avenue, Tucson, AZ 85719;

najita@noao.edu

---

[1] Visiting Astronomer at the Infrared Telescope Facility which is operated by the University of Hawaii under contract to the National Aeronautics and Space Administration

**ABSTRACT**

We report on the detection and analysis of hot ro-vibrational $H_2O$ emission from SVS-13, a young stellar object previously known to have strong CO overtone bandhead emission. Modeling of the high-resolution infrared spectrum shows that the $H_2O$ emission is characterized by temperatures ~ 1500 K, significantly lower than the temperatures that characterize the CO bandhead emission. The widths for the $H_2O$ lines are also found to be smaller than those for the CO lines. We construct a disk model for the emission that reproduces the CO and $H_2O$ spectrum. In this model, the $H_2O$ lines originate at somewhat larger disk radii ($\leq 0.3$ AU) than the CO overtone lines ($\leq 0.1$ AU). We find that the $H_2O$ abundance is about a factor of 10 lower than the calculated chemical equilibrium abundance.

Large, approximately transonic, local line broadening is required to fit the profile of the CO bandhead. If this velocity dispersion is identified with turbulence, it is of significant interest regarding the transport of angular momentum in disks. Large local broadening is also required in modeling CO overtone emission from other young stellar objects, suggesting that large turbulent velocities may be characteristic of the upper atmospheres of the inner disks of young stars.

Subject Headings: accretion, accretion disks — stars: pre-main-sequence — stars: formation — planetary systems: protoplanetary disks — circumstellar matter — infrared: stars

1. INTRODUCTION

The presence of hot molecular gas in the circumstellar environment of a number of young stellar objects (YSOs) has been established through observations of first overtone CO bandhead emission (Scoville et al. 1983; Thompson 1985; Geballe & Persson 1987; Carr 1989). The CO overtone emission arises from gas with temperatures ~ 3000 K and densities ≥ $10^{10}$ cm$^{-3}$. High-resolution spectroscopy has shown that the shape of the bandheads in most objects is consistent with emission from a rotating disk (Carr et al. 1993; Chandler, Carlstrom, & Scoville 1995; Najita et al. 1996). In lower mass YSOs, the emission must originate within a few tenths of an AU from the star. The emission lines are likely to be produced in a temperature inversion in the atmosphere of the accretion disk (Calvet et al. 1991; Najita et al. 1996).

More recent work has concentrated on the fundamental bands of CO (Carr, Mathieu & Najita 2001; Najita et al. 2003; Brittain et al. 2003). Emission in the fundamental bands is detected in most classical T Tarui stars (CTTS), unlike the situation for the CO overtone bands, which are detected in only a few percent of T Tauri stars. This difference in detection frequency can be attributed to the much higher A-values of the fundamental transitions and the lower energy of the v=1 level, which allow far smaller column densities and lower excitation gas to be detected. The fundamental emission in CTTSs is characterized by typical temperatures of ~ 1000 K, and the lines are often optically thick. Najita et al. (2003) argue that the emission is most likely to arise in a disk. In contrast to the often strongly double-peaked profiles for the overtone emission, the fundamental profiles are normally centrally peaked. This implies that the emission must extend over a large range in radii, from ~ 0.04 AU to ≥ 1 AU based on the velocity widths. Emission from the atmosphere of an optically thick disk is suggested, as for the case of the overtone bands;

however, in spectroscopic binaries, where the inner disk is expected to be tidally disrupted, the emission may arise from residual gas in disk gaps.

Molecules other than CO are expected to exist at the temperatures and densities in the inner few AU of disks. Water should be very abundant in the gas phase at temperatures below the dissociation temperature (~ 2500 K) and above the water-ice sublimation temperature (~ 150 K). In the absence of grains, $H_2O$ will be a major coolent and source of opacity. In general, $H_2O$ should be an excellent diagnostic of gas within the inner disk, due to its high abundance and its rich ro-vibrational and rotational spectrum in the near- to far-infrared, which can sample a large range in excitation conditions due to the wide range of energy levels and line strengths. While cool water vapor (< 500 K) is difficult to observe from the ground due to telluric absorption, hot water is easily studied. The detection of emission from hot $H_2O$ has already been reported from some YSOs known to have CO overtone emission (Najita et al. 2000). In this paper, we present and discuss in detail the detection of $H_2O$ emission from SVS-13, a YSO known to have prominent CO overtone bandhead emission.

## 2. OBSERVATIONS

High-resolution spectra of SVS-13 were obtained at the NASA Infrared Telescope Facility (IRTF) using CSHELL, the facility cryogenic infrared echelle spectrograph (Tokunaga et al. 1990; Greene et al. 1993), which is equipped with a 256x256 SBRC InSb array. The observations were made on 13 Nov. 1994 UT. CSHELL uses order sorting filters to isolate a single echelle order; for these observations, we used a custom narrow-band wedged filter, centered at 2.32 $\mu$m, that eliminates the channel fringing that was normally seen with the circular variable filters. Two overlapping grating settings centered near 2.293 $\mu$m were obtained. The 1" slit was used, giving a 5-pixel resolution R=23,000. For flux calibration, a short exposure of both SVS-13 and HD 18881 (K = 7.140) were obtained through a 4" wide slit. In the normal observing procedure, the telescope is nodded between integrations to place the star at two different positions along the slit, and these images are differenced during data reduction to remove dark and background. The extracted spectra of SVS-13 were divided by those of a hot star observed at the same airmass in order to correct for telluric absorption features. The telluric features in the hot star spectrum were used for wavelength calibration. The resulting telluric corrected and flux calibrated spectrum is shown in Figure 1.

A low-resolution (R = 200) spectrum of SVS-13 was obtained at the IRTF using SpeX (Rayner et al. 2003) with a prism covering the 0.8–2.5 $\mu$m spectral region. The observations were made on 01 Nov. 2001 UT. HD 19600 (spectral type AOV) was observed as a standard star. The atmospheric absorption and instrumental throughput were obtained by the method described by Vacca et al. (2003). The resulting spectrum of SVS-13 is shown in Figure 2(a).

## 3. RESULTS

The CSHELL spectrum in Figure 1 shows prominent emission from the CO v=2–0 bandhead as was first reported by Carr (1989) and subsequently observed at high-resolution by Carr & Tokunaga (1992) and Chandler et al. (1993). Significant spectral structure is also observed at wavelengths blueward of the CO bandhead ($\lambda$ < 2.2935 $\mu$m). The strongest of these emission features were found to correspond in wavelength with absorption lines of $H_2O$ that have been identified by Zobov et al. (2000) in sunspot spectra. The line assignments of Zobov et al. are marked in Figure 1. The $H_2O$ emission also appears to contribute to the blue wing of the CO bandhead.

Given the identification of the emission lines with high rotational transitions that are also observed in sunspot spectra, the water emission must originate from hot gas.

The presence of $H_2O$ emission is corroborated by the low-resolution SpeX spectrum. In Figure 2(b), we show the spectrum of SVS-13 dereddened in order to flatten and clearly show the entire 1.0–2.5 $\mu$m spectrum. Broad emission features are seen which correspond to the $H_2O$ bands at 1.4, 1.9 and 2.7 $\mu$m (the latter is blended with CO). These same bands are observed, in absorption, in the spectra of late-type stars.

## 4. MODELING

We have modeled the high-resolution spectrum of SVS-13 in order to determine the properties of the hot $H_2O$ gas and understand the relation between the $H_2O$ and CO overtone emission. We first describe the construction of a $H_2O$ line list for this purpose, using a combination of theoretical line lists and empirical modeling. We then model the $H_2O$ and CO spectrum, first as emission from gas in a uniform slab, and then as emission from a Keplerian disk.

### 4.1 $H_2O$ Linelist

The main difficulty encountered in modeling the spectrum of hot water is the availability of a complete and accurate $H_2O$ line list. One of the more extensive line lists suitable for our purpose is the theoretical list calculated by Partridge & Schwenke (1997, hereafter PS), which has been used in the most recent generation of atmospheric models for M dwarfs and brown dwarfs (Allard, Hauschildt & Schweitzer 2000; Allard, Hauschildt & Schwenke 2000; Allard et al. 2001). However, neither the positions nor strengths for individual lines in the PS list are accurate enough for synthesis of high-resolution spectra. Therefore, we took a semi-empirical approach, beginning with the PS list and modeling the $H_2O$ lines in a solar sunspot spectrum.

We used a recent version of the stellar synthesis program MOOG (Sneden 1973) and the infrared sunspot spectrum of Wallace & Livingston (1992). The MOOG code was modified to handle the triatomic molecule $H_2O$, using the partition function from Irwin (1988) corrected for the different treatment of the nuclear spin statistical weight. A comparison of the Irwin partition function with that calculated from the PS energy levels showed that the Irwin fit is accurate to better than 1 % down to a temperature of 500 K.

The PS list contains about 450,000 lines in the 2.288–2.297 $\mu$m wavelength region covered by our high-resolution spectrum of SVS-13. In order to make the list manageable for spectral modeling, we restricted the list to lines with strength greater than 1 % of the strongest line in the region (at 22918 Å) at a temperature of 3000 K. This leaves about 350 lines. We also produced lists with cutoffs of 0.1 % and 0.01 % of the strongest line, and lists using a 1 % cutoff at 1000 and 2000 K, in order to investigate the effects of the choice of cutoff threshold and temperature. We next took the line assignments for the solar sunspot spectrum and a laboratory emission spectrum by Zobov et al. (2000) and substituted the observed wavelengths for the corresponding transitions in the PS list. While these line assignments cover very few of the $H_2O$ lines in the sunspot spectrum, many of the strongest absorption lines in our wavelength region appear to be identified.

The adjustment of oscillator strengths for individual water transitions is the next issue to address. One approach is to match synthetic spectra to observed high-resolution spectra of a M dwarf or sunspot. We ruled out M dwarfs due to the lack of well determined abundances and uncertainty in the effective temperatures. One difficulty with sunspots is deciding on the atmospheric structure that accurately represents the observed sunspot. We used the NextGen model structures (Hauschildt, Allard, & Baron 1999) with solar metallicity and log g = 4.5. An abundance analysis of the CO overtone lines was used to determine the effective temperature that best corresponds to the Wallace & Livingston sunspot spectrum. An effective temperature of 3900 K gave a carbon abundance (log ε(C) = 8.64, using the notation log ε(C) = log (C/H) + 12.0) that was roughly independent of line excitation potential and close to the solar photospheric value. Even if this model is not a precise match to the observed sunspot, it should be adequate enough to adjust the *relative* oscillator strengths of $H_2O$ lines. However, this approach may not be sufficient to set the *absolute* oscillator strengths of the lines, because the $H_2O$ abundance is highly sensitive to the temperature structure.

To calibrate the absolute oscillator strengths, we used experimental line strengths taken from the HITRAN database (Rothman et al. 1998). Because none of the identified $H_2O$ lines in the wavelength region of the SVS-13 spectrum are in the HITRAN list, we used a wavelength region near 23110 Å, where several lines assigned by Zobov et al. (2000) appear in HITRAN. The gf-values for these lines, converted from the HITRAN line strengths, were substituted into the PS list. We then produced synthetic spectra using the above model and C abundance and adjusted the O abundance to give the best fit to four clean HITRAN $H_2O$ lines (the solar value log ε(O) = 8.93 +/- 0.04 gave the best fit). All four of these lines are from the 001-000 band, and the HITRAN based gf-values were 1.17 to 1.35 times greater than the theoretical values in the PS list. We next synthesized sunspot spectra for the region covered by the SVS-13 spectrum, with the C and O abundances fixed, and adjusted the gf-values of the lines identified by Zobov et al. to fit the observed line strengths. These adjustments required an increase of the theoretical gf-values by factors ranging from 1.1 to 2.7; the changes in gf-values for only the 001-000 band lines were 1.1 to 1.5 times, similar to the ratio of the HITRAN to PS gf-values in the 23110 Å region. With this procedure, we believe that the absolute gf-values for the identified lines are good to 25 % and the relative gf-values good to 10 %.

Figure 3 shows the observed and synthetic sunspot spectra for the wavelength region blueward of the CO bandhead. The $H_2O$ line at 22913 Å could not be fit because of strong telluric absorption; the gf-value for this line was set to 1.3 times the PS theoretical value, which is the median ratio of empirical to theoretical gf-values for the other 001-000 lines. There are some prominent lines in the sunspot spectrum (22882, 22892 and 22908 Å) that are not accounted for. However, emission centered at these wavelengths is clearly not present in the SVS-13 spectrum. These absorption lines could be due to a molecule other than $H_2O$ or to higher excitation $H_2O$ transitions that are not populated in the (presumably cooler) circumstellar gas of SVS-13. The theoretical line list also leaves significant opacity due to moderate strength lines unaccounted for (e.g., near 22887 and 22929 Å). The inadequacy of current theoretical $H_2O$ line lists is a general problem which also reveals itself in the spectral modeling of late-type stars at low resolution (Allard et al. 2000b; Leggett et al. 2000, 2001; Jones et al. 2002). In order to check that our reduced line list is not missing important opacity, we ran the same model using our line lists with cutoffs of 0.1% and 0.01 % of the strongest line. The 0.1% cutoff list introduces a veil of very weak lines that depresses the continuum by 0.5 % but does not produce any additional spectral features. Use of the line list with a 0.01 % cutoff produced no further change. The line lists that were constructed with cutoffs based on temperatures of 1000 K and 2000 K gave similar results, except that the lower temperature line lists missed some weak high-excitation lines.

### 4.2 Slab Model

We begin with a simple model with gas of constant temperature, column density $\Sigma$, and number density. The gas is assumed to be in chemical equilibrium, and the computation of the chemical equilibrium abundances, which depends only on the local number density and temperature, included all important C and O bearing molecules. In addition, LTE was assumed.

Continuum opacity was not included. The gas has a local line profile composed of both a thermal and non-thermal component. The non-thermal component is characterized by a velocity dispersion $\xi$ ($\Delta V_{FWHM} = 2.355\cdot\xi$), with either a Gaussian or Lorentzian profile. A Gaussian macro-broadening is also included to represent bulk gas motions on scales larger than the photon mean free path. The instrumental broadening is a Gaussian with a 13 km s$^{-1}$ FWHM velocity. For this modeling, we sought to match the relative shape and line ratios of the emission; hence, the data were normalized to a continuum level of one, and the model spectra were scaled to match the integrated emission flux of either the CO or H$_2$O.

First, we fit the CO 2–0 bandhead using a list with only CO lines. The shape of the bandhead is very much affected by the overlap of the CO lines and the line optical depth. We found that solutions where the CO lines are optically thin cannot fit the shape of the bandhead, regardless of the temperature, because the model is too sharply peaked at the bandhead (22938 Å). Models where the local line profile includes only thermal broadening do not fit for any combination of temperature and column density; such a model also results in a bandhead that is too sharply peaked with too little flux at wavelengths longer than 22945 Å. The necessity for superthermal local linewidths was also found when fitting the CO bandhead emission from other YSOs (Najita et al. 1996); the line broadening will be discussed in § 5. Acceptable fits could be found over the temperature range of 1500 – 3500 K, which reflects the fact that the CO 2–0 bandhead, by itself, is more sensitive to optical depth and line broadening than to temperature. As an example of a good fit, at 2000 K, $\Sigma = 8.0$ g cm$^{-2}$ and $\xi = 10$ km s$^{-1}$ (Gaussian), with a macro-broadening of 38 km s$^{-1}$ (FWHM). These parameters give $\tau = 0.5$ in the 2–0 R(51) line. Alternately, a similar fit can be obtained using a Lorentzian non-thermal profile with $\xi = 2.8$ km s$^{-1}$ and $\Sigma$ = 12 g cm$^{-2}$.  In general, different shapes for the local line profile give similar results because the bulk gas motions dominate the total line broadening.

Next, we fit the H$_2$O lines blueward of the bandhead using only H$_2$O lines, but we added to the model the previous 2000 K CO model in order to account for the contribution of the CO bandhead wing near 22933 Å.  The density was set sufficiently high (10$^{12}$ cm$^{-3}$) that H$_2$O has its maximum chemical equilibrium abundance.  The H$_2$O lines with calibrated line strengths have a range in excitation potential of 0.3 to 0.6 eV, and hence provide some diagnostic of the gas temperature.  A temperature of ~ 1500 K worked best, while T = 1000 K gave an inadequate fit and 2000 K was definitely too hot.  We found that different combinations of $\xi$ and $\Sigma$ gave similar results because overlap of the main H$_2$O lines is not important, unlike the case for the CO bandhead.  In all cases, optically thick H$_2$O lines provided a better fit than optically thin lines.  For example, a reasonable fit at 1500 K is achieved with $\Sigma$ = 9.0 g cm$^{-2}$ (giving $\tau$ = 6.0 in the 22918 Å line), $\xi$ = 10 km s$^{-1}$ (Gaussian), and a macro-broadening of 21 km s$^{-1}$.  This model is shown in Figure 4a (smooth solid-line) overplotted on the observed spectrum.  The apparent wing to the CO bandhead at 22931 Å is fit well with the inclusion of the H$_2$O emission.  This wing could not be fit with the CO models in Chandler et al. (1995) and Carr & Tokunaga (1992).  Note that the line broadening for the H$_2$O lines is significantly less than for the CO bandhead.  This is illustrated by the dotted-line in Figure 4a, which is the same H$_2$O model but with the 38 km s$^{-1}$ macro-broadening that was required for CO.

The best temperature of ~ 1500 K is much lower than the 3000 K used to determine the cutoff threshold in our main line list.  Therefore, we ran one test case for the above 1500 K model using the line list constructed for 1000 K.  Some additional low-excitation lines are present in the 1000 K list, but these lines do not produce any significant difference in the model spectrum.

For a slab model, we find that there is no set of parameters that can simultaneously fit the $H_2O$ and CO emission. The first, and obvious, difficulty is that the $H_2O$ lines are narrower than the CO lines (Fig. 4a). The second problem lies in the gas temperature. The $H_2O$ emission indicates a temperature ~ 1500 K. This temperature is at the low end of the range that can fit the CO 2–0 bandhead; however, a temperature this low for the CO emission can be ruled out by the presence of emission from higher vibrational bandheads (e.g., v=5–3), which requires gas temperatures ≥ 2500 K (Carr 1989). In addition, too much $H_2O$ emission is produced at 1500 K relative to CO, and the $H_2O$ column density must be reduced substantially to match the relative flux. This can be done by reducing the number density and hence the $H_2O$ chemical equilibrium abundance. However, the $H_2O$ lines then become optically thin ($\tau_{22918} = 0.27$), and the resulting fit to the $H_2O$ line ratios is poorer (see solid line in Fig. 4b). Higher gas temperatures do not improve the situation. Figure 4(b) also shows a model (dotted line) with parameters that fit the CO bandhead for T = 2000 K, where the number density is again adjusted to fit the relative $H_2O$ to CO flux. The optical depth is still too low ($\tau_{22918} = 0.36$), and while the higher temperature helps with the 22931 Å feature, the other line ratios are worse. We conclude that a single slab of uniform gas cannot produce both the $H_2O$ and CO emission and that a range of temperatures and linewidths is required.

### 4.3 Disk Model

Because a disk model works well for the CO bandhead emission in many YSOs (Carr et al. 1993; Chandler et al. 1995; Najita et al. 1996), this is a logical model to explore for the $H_2O$ emission. $H_2O$ has a lower dissociation temperature than CO, therefore, a disk with an outwardly decreasing temperature gradient will naturally give $H_2O$ emission lines with smaller linewidths and lower excitation temperatures, compared to the overtone CO emission.

In this LTE disk model (see Najita et al. 1996; Carr et al. 2001), the molecular emission arises from a line emitting layer that is uniform in the vertical direction and whose radial variation of temperature and surface density $\Sigma$ are specified by power laws. This line emitting layer may be a temperature inversion in the upper atmosphere of an optically thick disk, or it could correspond to the entire surface density of an optically thin portion of the disk. The chemical equilibrium abundances are calculated as in the slab model, but in this case the average number density comes from the surface density and the thickness of the layer, taken as the pressure scale height in hydrostatic equilibrium. The local line broadening is treated the same as in the slab model, but now all macro-broadening is determined by the Keplerian rotation of the disk. The other quantities that must be specified are v sin i at the inner emission radius, the inner and outer radius of the emission, and the mass of the star; the inclination follows from these parameters. Because we first concentrate on fitting the shape of the CO bandhead and the width and ratios of the $H_2O$ lines, the relevant parameters are the temperature and surface density laws, v sin i, and the *ratio* of the outer to inner *emission* radii, $R_{out}/R_{in}$. The stellar mass has only a minor effect on the model; we take 2 $M_\odot$ based on the bolometric luminosity of SVS-13. The absolute value of the radius is not critical except when fitting the absolute observed flux, in which case the distance and extinction are also required. We will consider the absolute flux after first exploring the parameters required to fit the spectral shape.

The results of the slab modeling provide a guide to the average conditions that are required to fit the data. Because of the many adjustable parameters in the disk model, unique solutions are not possible. Instead, we present a model that is illustrative of solutions that provide a good match to the observed spectrum. This model has the following parameters: $T = 4000 \cdot (r/R_{in})^{-0.4}$ K, $\Sigma = 12 \cdot (r/R_{in})^{-0.4}$ g cm$^{-2}$, $\xi = 11$ km s$^{-1}$, v sin i = 48 km s$^{-1}$, and $R_{out}/R_{in} = 22$. For a Lorentzian local profile, $\Sigma = 22 \cdot (r/R_{in})^{-0.4}$ g cm$^{-2}$ and $\xi = 3.2$ km s$^{-1}$ give an equivalent result. The exponents for the temperature and surface density power laws are arbitrary but plausible values. The values for $\Sigma$, $\xi$, and v sin i were set by fitting the shape of the CO bandhead, while $R_{out}/R_{in}$ was set to provide the correct average temperature for the H$_2$O gas. The temperature at the outer radius is 1160 K, and $\Sigma = 3.5$ g cm$^{-2}$.

A possible physical interpretation for an outer cutoff in emission at this radius and temperature is the presence of dust in the upper atmosphere. The temperature at the outer radius is very close to the grain vaporization temperature (Pollack et al. 1994) for the model number density of $5 \times 10^{11}$ cm$^{-3}$ at this radius. Using the Rosseland mean opacities from Henning & Stognienko (1996), the continuum optical depth due to dust in the layer would be $\tau = 20$. This is high enough to effectively suppress the molecular line emission.

Figure 5 shows this model when the flux is scaled to match the observed flux in the CO bandhead. The solid line in Figure 5 is the model spectrum using only CO lines. When the H$_2$O lines are added to the synthesis, it is found that the H$_2$O line flux, relative to CO, is far too large. The molecular abundances in chemical equilibrium are set by the temperature and number density at each radius. It is not possible to independently vary the number density, as was done in the slab model, because the number density is set by the surface density, temperature, and pressure scale height for the gas layer. Instead, we allow the H$_2$O abundance to be decreased by an arbitrary amount below its chemical equilibrium value. The dashed line in Figure 5 shows such a model, where the H$_2$O abundance is scaled (by a factor of 0.14) to fit the wing to the CO bandhead.

In Figure 6 we show the *same* baseline disk model (solid line), but scaled instead to the flux under the H$_2$O lines between 22913 and 22922 Å. This model gives a reasonable match to the H$_2$O linewidth, temperature, and optical depth, though a somewhat better fit (similar to the slab model in Fig4a, solid-line) would result by increasing $\Sigma$ by a factor of two. However, when scaled to the $H_2O$ flux, the model CO emission is now far too weak, as is seen by the low CO flux at 22937 Å (solid line). The dotted line in Figure 6 shows the synthetic spectrum when the model is scaled to the CO bandhead and the $H_2O$ abundance is reduced. This model is identical to the dashed line in Figure 5, expect that the $H_2O$ abundance is scaled to match the $H_2O$ flux between 22913 and 22922 Å rather than the wing at 22931 Å. This result is similar to the problem encountered when reducing the $H_2O$ abundance in the slab model – decreasing the $H_2O$ column density to match the relative $H_2O$ and CO fluxes produces optically thin $H_2O$ lines that give a poorer match to the $H_2O$ line ratios. This may indicate that our disk model is too simple (see discussion in § 5). However, some caution must be exercised regarding our findings about the $H_2O$ optical depth (and excitation temperature) due to the relatively small number of $H_2O$ lines used in the analysis. More definitive conclusions would be provided by an analysis of a far larger number of lines.

Next, we examined the physical radius of the emission that is required to fit the absolute CO flux level in the spectrum. For the distance to SVS-13, we take 350 pc (Herbig & Jones 1983). The extinction is an important but not well determined factor. The infrared color-color plots in Aspin & Sandell (1997) suggest an extinction in the range of $A_V = 15 - 25$, or $A_K \sim 2$. If we set $R_{in}$ = 3.0 $R_\odot$, for example, then $A_K = 2.05$ ($A_V = 19$) is required to match the CO bandhead flux. The inclination is i = 8° for this combination of radius, stellar mass and v sin i. The $H_2O$ emission in this model would extend out to ~ 0.3 AU. If the extinction is $A_V = 25$ ($A_K = 2.7$), near the upper end of this range, then a radius of $R_{in} = 4.0$ $R_\odot$ is needed, and the emission would extend to ~ 0.4 AU. $R_{in}$ could be larger by a factor of a couple if the maximum gas temperature in the model is decreased. Regardless of the exact values for the input parameters, the observed CO flux requires the emission region to be rather small, with the inner edge within a few stellar radii of the star.

Figure 7 shows where the CO and $H_2O$ emission are produced in the baseline disk model as a function of radius. The CO and $H_2O$ emission come from overlapping but largely different parts of the disk. The inner radii are determined by molecular dissociation. The power weighted average temperatures for the CO and $H_2O$ emission are 2770 and 1680 K, respectively. The typical values of v sin i are 31 and 17 km s$^{-1}$ for CO and $H_2O$, respectively.

## 5.0 DISCUSSION

We have shown, using both low and high-resolution spectra, that emission from hot $H_2O$ is present in SVS-13. After constructing a $H_2O$ line list suitable for modeling high-resolution spectra, we modeled both the CO bandhead and $H_2O$ emission. The velocity widths of the $H_2O$ emission lines are narrower than those for the CO emission, and the $H_2O$ excitation temperature of ~ 1500 K is significantly lower than the temperature ($\geq$ 2500 K) required to produce CO emission from high vibrational levels. We constructed a model of emission from a Keplerian accretion disk that naturally provides the correct temperatures and linewidths for both the CO and $H_2O$ emission and gives a reasonable match to the observed shape of the CO and $H_2O$ spectra. In this disk model, the peak of the $H_2O$ emission lies at a larger radius than the CO emission (Fig. 7), extending to ~ 0.3 AU from the star, compared to ~ 0.1 AU for the CO overtone emission.

In the baseline disk model, the molecular abundances at each disk radius were determined assuming chemical equilibrium. However, we found that the $H_2O$ abundance must be substantially reduced below this value, by a factor of ten, in order to match the $H_2O$ emission flux relative to CO. Strong incident radiation on the disk surface may be responsible for producing molecular abundances that differ from expected equilibrium values. Photodissociation from UV photons is likely to be important near the disk surface, and strong Ly $\alpha$ emission could preferentially destroy some molecules, such as $H_2O$, compared to CO (Bergin et al. 2003). In addition, the energetic X-rays emitted by T Tauri stars are a major source of ionization in the disk atmosphere (Glassgold, Najita & Igea 1997; Igea & Glassgold 1999) that directly affect the chemical abundances (Aikawa & Herbst 1999, 2001; Glassgold & Najita 2001).

Preliminary modeling by Glassgold & Najita (2001) also shows that X-ray induced heating and chemistry in the inner disk can produce a temperature inversion in the gas and a strong vertical chemical structure. Such a vertical structure would likely alter interpretations based on the simplistic uniform structure in our disk model. For example, the apparent low abundance of $H_2O$ could arise if CO is present over a larger vertical column than $H_2O$, with $H_2O$ found at larger depths and lower temperatures in the disk atmosphere. In order to sort out the thermal-chemical structure of the inner disk, and to understand the processes that produce that structure, physical models that calculate both the radial and vertical structure of the atmosphere will be needed.

Another alternative, that could mimic a reduced abundance, is a continuum opacity that is just sufficient to reduce, but not totally suppress, the line emission. Grains are not expected in thermodynamic equilibrium (Pollack et al. 1994) for the temperatures and densities in the $H_2O$ emission layer. However, if small amounts of dust survived or were mixed into the upper atmosphere from lower disk layers, then sufficient opacity could be produced. This suggests that the details of grain evaporation, condensation, and transport are potentially important considerations for the measurement of molecular abundances.

An interesting result of the spectral modeling is the large local line broadening that is required to fit the CO bandhead. The near overlap of CO transitions near the v=2-0 bandhead is the key that allows a separation of the local broadening from the macro-broadening. For the baseline disk model with a local Gaussian profile, the non-thermal velocity dispersion was 11 km s$^{-1}$. Acceptable fits were obtained for $\xi$ between 7 and 15 km s$^{-1}$ (with changes in $\Sigma$ and v sin i). These values are far larger than the thermal dispersion of 0.9 km s$^{-1}$ for CO at 2500 K and larger than the sound speed of 5.2 km s$^{-1}$. The baseline model with a Lorentzian profile required $\xi = 3.2$ km s$^{-1}$. While the core of this Lorentzian profile is subsonic, supersonic motions still occur in the line wings.

Evidence for supersonic line broadening has also been presented for the accretion disks of cataclysmic variable stars. Horne et al. (1994) found that a velocity dispersion of ~ Mach 6 was required to fit the veil of Fe II absorption features that were attributed to intervening disk material in the eclipsing system OY Car. Supersonic broadening was also found for several cataclysmic variable star disks by Marsh & Dhillon (1997). Their method was not dissimilar to ours, in that it relied upon the profile overlap of two close atomic lines; however, in their case the large line widths could potentially be produced by Stark broadening.

It is tempting to identify the local line broadening with turbulence in a disk. Turbulent motions in disks are believed to be closely linked to a long-standing difficulty in the understanding of accretion disk physics, namely the physical process responsible for angular momentum transport in disks. Several possibilities have been suggested in the context of T Tauri disks. One attractive suggestion is the magnetorotational instability (Balbus & Hawley 1991; Stone et al. 2000), in which a weak magnetic field in a conducting, differentially rotating fluid renders it unstable to turbulence. More recently, Klahr & Bodenheimer (2003) have proposed that the global baroclinic instability, a purely hydrodynamic instability, is a source of turbulence that leads to angular momentum transport in disks. Either of these processes could operate in the region probed by our observations (r < 0.4 AU); in particular, at the temperatures indicated by the modeling (T > 1200 K), thermal ionization is sufficient for the magnetorotational instability to be operative if a weak magnetic field is present. Both instabilities are expected to produce turbulent velocities that are less than or comparable to the sound speed. Simulations of these instabilities also show that the velocity dispersion increases with height in the disk, eventually reaching the sound speed (Klahr & Bodenheimer 2003; Miller & Stone 2000) in the upper disk atmosphere. The upper atmosphere of the disk is the region where we believe the CO and $H_2O$ emission lines originate.

In our modeling, the fits with Gaussian local profiles would imply supersonic turbulence of ~ Mach 2. Models using a Lorentzian profile, however, show that it is possible to fit the CO bandhead using a profile that has a Gaussian core with a subsonic width, but stronger than Gaussian wings at high velocities. Hence, a non-Gaussian profile would allow the bulk of the motions to be transonic or subsonic and consistent with the velocities expected for turbulence in protoplanetary disks. In addition, non-Gaussian line profiles might be expected, given that a characteristic of turbulence seen in both laboratory experiments and numerical simulations are non-Gaussian wings in the probability density functions (PDFs) of the velocity (Frisch 1995; references cited in Falgarone et al. 1994). Indeed, it has been argued that evidence of turbulence in the interstellar medium is provided by the non-Gaussian high-velocity wings of molecular line profiles (Falgarone & Phillips 1990; Falgarone et al. 1994) , and the non-Gaussian wings in the PDFs of centroid velocities or centroid velocity increments (Miesch & Scalo 1995; Lis et al. 1996, 1998; Miesch, Scalo & Bally 1999).

If the local, non-thermal, line broadening is due to disk turbulence, then analyses of other stars, and measurements of molecular transitions that probe different temperatures and radii, would be of great interest. Modeling of CO overtone emission by Najita et al. (1996) in two other YSOs also required large local velocity dispersions, though not quite as large as the broadening we require for SVS-13. If large velocity dispersions are characteristic of the molecular line emitting regions of the inner disk, this may help to explain why centrally peaked, rather than double-peaked, profiles are commonly observed in the CO fundamental emission lines in T Tauri stars (Najita et al. 2003). Modeling of these CO fundamental lines can provide the magnitude of the local velocity dispersion. Our current data for SVS-13, unfortunately, do not allow an independent determination of $\xi$ for $H_2O$, but this could be accomplished if some $H_2O$ lines with the correct velocity separation can be identified and analyzed. In any case, further high-resolution spectroscopy and radiative transfer modeling of disk emission lines could provide observational evidence for the turbulence responsible for angular momentum transport in disks and supply information on both the magnitude and character of the turbulent motions (see, e.g., Horne 1995 on the possibility of detecting anisotropic turbulence).

A number of other YSOs with CO overtone emission show emission from hot $H_2O$ (Najita et al. 2000). However, the current data are insufficient to say how common hot $H_2O$ emission is among young stars. CO fundamental emission is commonly observed in T Tauri stars (Najita et al. 2003), and the $H_2O$ excitation temperature is closer to the typical temperature of the CO fundamental emission than to that of the overtone emission, which suggests that hot $H_2O$ emission could be common. On the other hand, the total column densities deduced for the $H_2O$ emission ($\Sigma \sim 1\text{-}10$ g cm$^{-2}$) are similar to those required for CO overtone emission and larger than typical of fundamental emission ($\Sigma < 0.1$ g cm$^{-2}$); hence, the $H_2O$ transitions we have observed in SVS-13 may be too weak to measure in a typical T Tauri star. The stronger, lower-excitation, $H_2O$ transitions in the mid-infrared may be a more promising prospect to study water in protoplanetary disks. Because the rich spectrum of $H_2O$ in the infrared could make water emission an effective probe of disks, further observations in both the near- and mid-infrared to determine the occurrence and characteristics of $H_2O$ emission are warranted.

J. S. C. acknowledges support from the NASA Origins of Solar Systems program and the Office of Naval Research. A. Tokunaga acknowledges the support of NASA Cooperative Agreement NCC 5-538. We wish to thank D. Schwenke for providing tapes of the PS water line list.

**FIGURE CAPTIONS**

Figure 1. High-resolution CSHELL spectrum of SVS-13 in the region of the v=2–0 CO bandhead. The $H_2O$ line assignments for the sunspot spectrum from Zobov et al. (2000) are labeled.

Figure 2. (a) Low-resolution 0.8–2.5 $\mu$m SpeX spectrum of SVS-13. (b) The same spectrum dereddened (for $A_K = 0.8$) in order to emphasize the $H_2O$ emission bands, whose positions are indicated by the bold horizontal lines. The positions of the first overtone CO bandheads are marked by the vertical lines.

Figure 3. Observed sunspot spectrum (thick line) from Wallace & Livingston (1992) and synthetic spectrum (thin line) calculated as described in the text. The positions of assigned $H_2O$ lines shown in Figure 1 are indicated. A ? labels the strongest unidentified absorption lines. Also labeled are a line of HF and a Zeeman-split Ti I line (neither are included in the synthesis). Wavelength regions with the strongest telluric absorption are not plotted in the observed spectrum.

Figure 4. (a) Slab model (smooth solid line) that provides a good fit to the $H_2O$ emission lines. The parameters for the model are T = 1500 K, $\Sigma = 9.0$ g cm$^{-2}$, $\xi = 10$ km s$^{-1}$, and a Gaussian macro-broadening of 21 km s$^{-1}$. The $H_2O$ lines are optically thick. The dashed line shows the contribution due to CO alone. The dotted line is the same model but with the macro-broadening of 38 km s$^{-1}$ that best fits the CO bandhead, illustrating the different broadenings for the CO and $H_2O$ lines. (b) Models that fit the CO bandhead compared to the observed $H_2O$ lines, after adjusting the number density to change the $H_2O$ chemical equilibrium abundance and match the relative CO to $H_2O$ fluxes. The solid line is for T = 1500 K, $\Sigma = 8.0$ g cm$^{-2}$, $\xi = 10$ km s$^{-1}$, and a Gaussian

macro-broadening of 38 km s$^{-1}$. The H$_2$O lines are now optically thin, and the fit is not as good as the solid-line in Fig. 4a. The dotted line is for T = 2000 K and $\Sigma$ = 10.0 g cm$^{-2}$.

Figure 5. Baseline disk model described in the text compared to the CO 2–0 bandhead in SVS-13 (histogram). The solid line is the disk model with only CO lines included, scaled to the flux under the CO bandhead. The dashed-line includes the H$_2$O lines, with the H$_2$O abundance decreased by 0.14 from its chemical equilibrium value to match the wing at 22933 Å.

Figure 6. Disk models and observed H$_2$O lines in SVS-13 (histogram) shortward of the CO bandhead. The smooth solid line is the baseline disk model scaled to the flux under the H$_2$O lines, where the H$_2$O abundance is set to its chemical equilibrium value. The dotted line is the baseline model scaled to the flux under the CO bandhead, but the H$_2$O abundance is decreased from its chemical equilibrium value to fit the average H$_2$O flux. With the lower abundance, the H$_2$O lines are optically thin and the fit to the spectrum is not quite as good.

Figure 7. Normalized power per unit radius for CO (solid line) and H$_2$O (dotted line) emission as a function of radius, for the baseline disk model (see text) with $R_{in}$ = 3.0 $R_\odot$, and $A_K$ = 2.05. The dashed line shows the gas temperature of the line emitting layer as a function of radius. The upper axis shows the value of v sin i corresponding to the radius on the lower axis.